Garance BENOIT
Ca'Foscari University
Venice, feb. 2024

# MATTER AND COSMOGENESIS
# IN KANT'S THEORY OF THE HEAVENS

As someone who works on post-Copernican astronomy, I wanted to talk to you today about matter from a cosmogonic perspective.

So why is it appropriate for a meeting on matter to talk about cosmogony, a theory on the formation of the world, and first on the organisation of celestial phenomena? In my opinion, there are three reasons for this:

The first is that cosmogony is the the place or the domain where we can attest to the productivity of matter. In this sense, the cosmogonic enterprise shows us that matter is not just a substrate, that we don't necessarily need Platonic forms to make a world. These stories even reveal that matter has its own organisational principle. It is a question of starting from a state that is almost not organised (I'll come back to this 'almost not') and arriving at a diversity of phenomena. This is so true that Cartesian and post-Cartesian theories played a paradigmatic role in the thinking of the Enlightenment in the formation of human societies (Rousseau, in his second discourse, claims to be inspired by Descartes' method of cosmogony), as well as in the formation of knowledge from sensations (as in the Condillac's work).

The second point concerns this « almost not » organised. The remarkable thing from the Ancients to the eighteenth century, is that whenever the heavens are mentioned, matter is organised into elements. These elements reveal both the grain of phenomena and their total organisation.

Thirdly and finally, this theory of the elements is often indexed to the functioning of a sense (touch in Aristotle, vision in Descartes), so that for the observer the aim of cosmogony is to organise the material phenomena of the world. It aims at the appearance, the phenomenality of the material world as a whole. In short, it is a theory of the perception of matter.

For all these reasons, by the beginning of the eighteenth century, cosmogony had practically become a genre among writers who multiplied these accounts in order to explain a certain organisation of the material world, starting from simple elements. With one notable exception: that of Newton, whose system dominated the physical sciences and astronomy, but whose rigorous mind had forbidden this type of overly speculative production.

Kant, who was a great admirer of Newton, set out to reconstruct a system compatible with the English physicist's. He did this in a very original text called *The General History of Nature and Theory of Heaven*. And it is of this pre-critical text that I would like to speak to you today.

## I The state of the question (the dichotomy)

Kant was aware of the theories of his time and found in some of their impasses the opportunity to formulate new hypotheses. From the very first lines of the book, he presents what may appear to be an antinomy between two ways of explaining the functioning of the system, both equally admissible but both equally inadequate:

On the one hand, when we observe the systematic character of our world (the Sun and its procession of planets), we are tempted to say that the same material cause orders or has ordered it: the planets all orbit in the same direction, in the same plane, and their distribution in space corresponds to variations of increasing density as we approach the star. But, on the other hand, if we examine the space in which these planets orbit, we find that it is completely empty, which prevents us from attributing a material cause to their movement.

In this twofold presentation, we can see the coordinates of a dilemma that had prevailed for a century between two cosmological models.

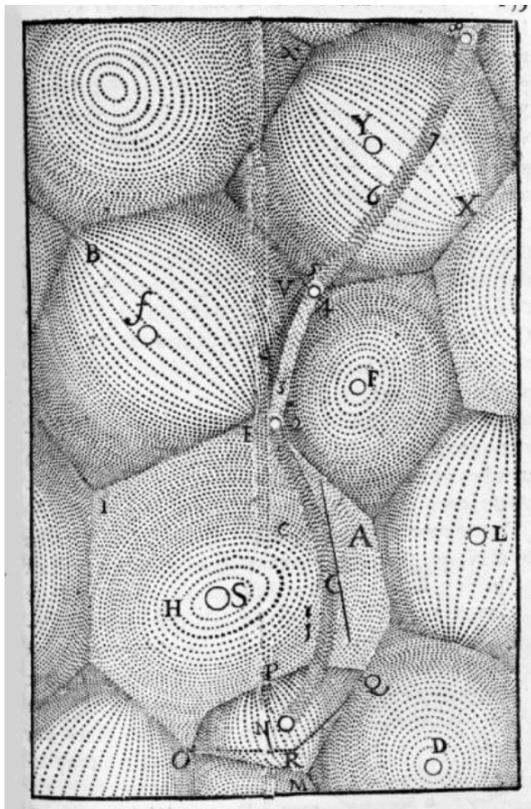

Descartes' theory of matter vortices in *Principia Philosophicae*, 1644.

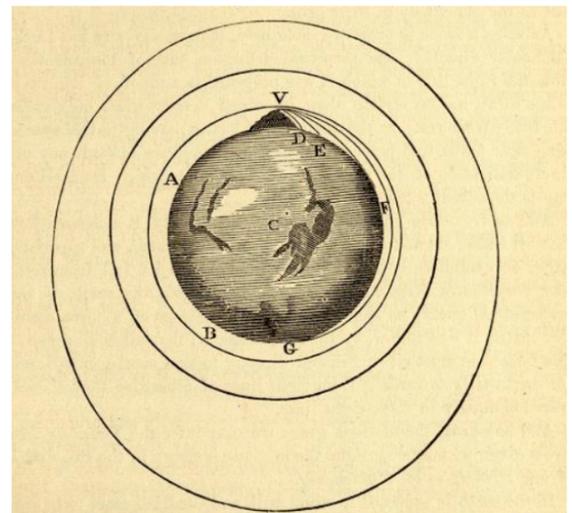

Newton's description of attraction (*Principia mathematica*, 1687).

In the third part of his Principles of Philosophy, Descartes developed a theory of the vortices of matter that carried the stars across a sky without empty space. This account had the advantage of explaining the emergence and continuity of the system's motion through an integral mechanistic materialism that did not require an appeal to abstract metaphysical entities or to a continuing intervention by the hand of God.

This theory had an undeniable advantage in terms of the material understanding of the system, but it also contained mathematical impasses which Newton did not fail to point out in his published Mathematical Principles. The details of the Cartesian vortices were incompatible with Kepler's laws, and they ignored the force of resistance of liquids, which should ultimately have stopped the vortices of matter (which is ironic, given that Descartes criticised Galileo for ignoring the force of resistance of air in his quantification of the fall of bodies).

At the time of Kant's writing, we were no longer in what in astronomy has come to be known as a situation of competing hypotheses, where two systems are equally valid from the point of view of phenomena, with no way of deciding between them. Kant and his whole century chose: Newton won the battle of ideas. But there is a well-known gap in the intelligence of Newtonian theory, which explains the persistence of vortices theories until the first half of the eighteenth century (like the one of Johannes Bernoulli). This lack is not to be found at the mathematical level, but in physics and natural philosophy, in the search for causality. Newton had forbidden himself to express an opinion on the cause of attraction in a well-known adage taken from the General Scolia of Book II of the *Principia mathematica*: "hypothesis non fingo" = I will not form a hypothesis on the cause of attraction. Even if, in reality, he spent the rest of his life searching for that cause in notebooks…

## II The Kantian enterprise

In order to overcome this opposition, and to give an account of material causality that could be reconciled with Newton's theory, Kant proposed an original history of the system. To do this, he took, as his starting point, new observations made by Maupertuis and Thomas Wright, and speculated on them to derive the idea of two of the most famous objects in our current cosmology: galactic formations and primitive nebulae.

The occasion for Kant's cosmogenetic enterprise was first a discovery made by Maupertuis in the *Discourse on the figure of the stars*. The purpose of this treatise was to present the main arguments in the opposition between Descartes and Newton, but he mentions in passing the observation of unknown "nebular formations" close to well-known stars, and observed thanks to progress in optics. Maupertuis thought they were massive stars, but Kant had reservations about this interpretation: if these nebulae were indeed stars, their luminous radiation should be proportional to their size, which is clearly not the case.

So Kant suggested instead that they were clusters of stars. And he compared this hypothesis with another proposed by Thomas Wright in his book *An Original Theory or New Hypothesis of the Universe*, based on his observations of the Milky Way. For him, the Milky Way should be interpreted as the distribution of a myriad of stars in a single plane . From this, Kant deduced that the spherical nebulae we observe in the sky are the total form of what he called « island-universes », galaxies of stars distributed in a circle on the same rotating plane, of which the Milky Way is the external mark that we see insofar as we are inside in one of them. What we really see is the slice of the disc that includes us.

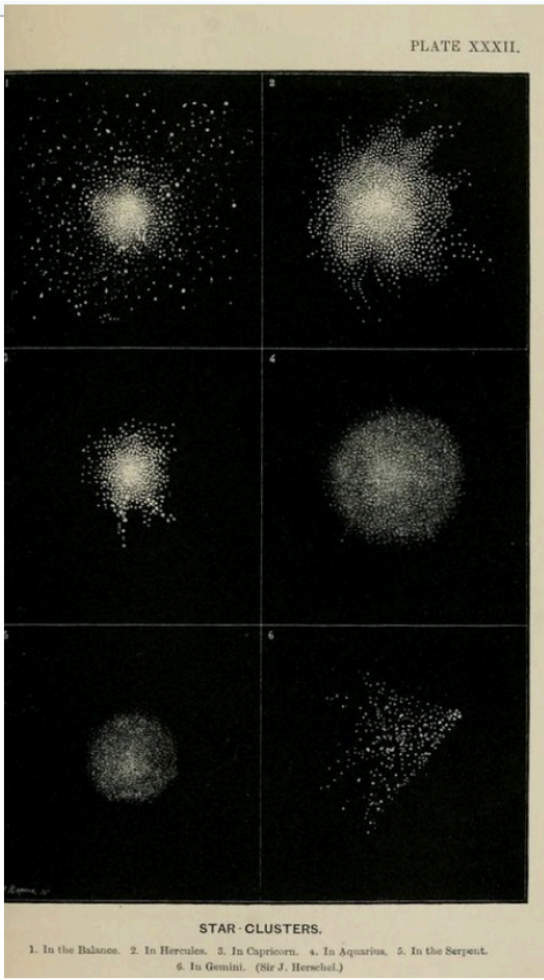

Star clusters, Herschel, *Treatise on astronomy* (1834).

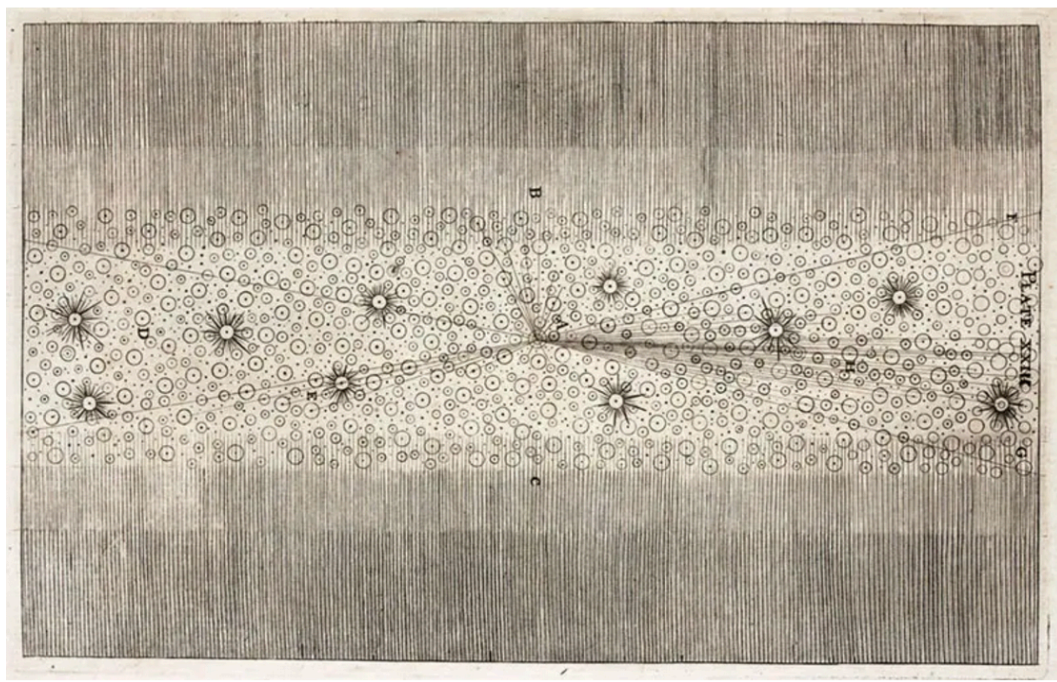

A section of the Milky Way, showing stars as arranged on a broad thin plane, an innovative idea of Thomas Wright, engraving in his *An Original Theory or New Hypothesis of the Universe*, 1750.

These are the first drawings of galaxies imagined by Thomas Wright and, a little later, the first transcriptions of observations of it by Herschel :

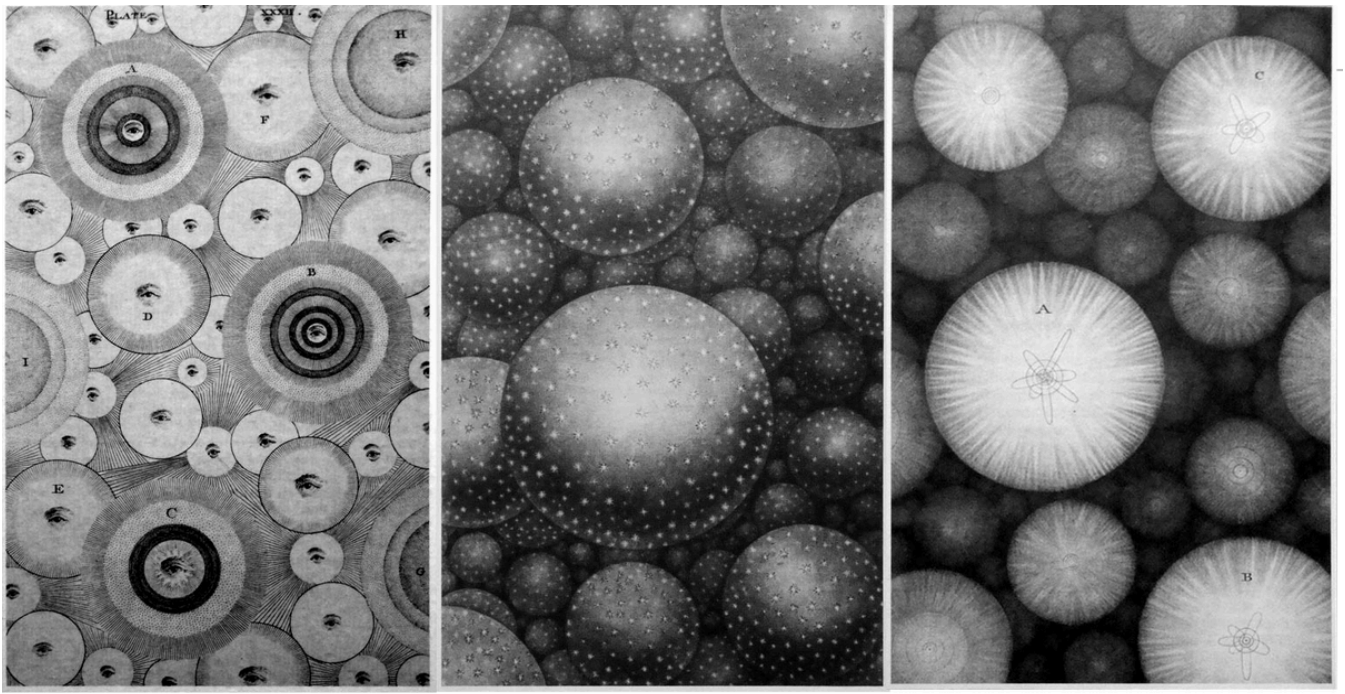

Depiction of Wright's theory of the existence of galaxies beyond our Milky Way in his *An Original Theory or New Hypothesis of the Universe*, 1750.

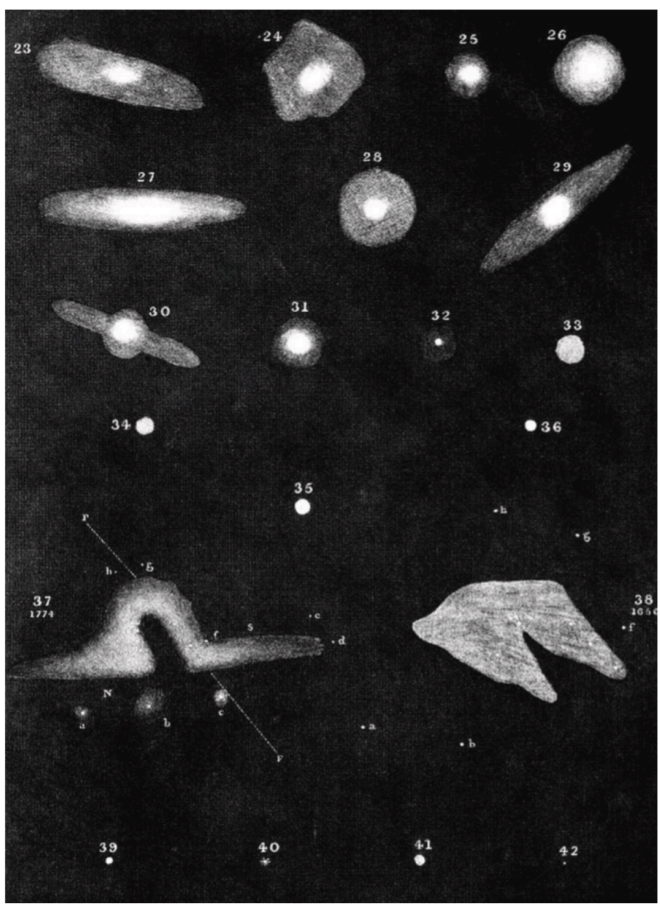

View of nebulae, Herschel, *Treatise on astronomy* (1834).

From there, and this is Kant's major contribution, he went on to imagine another type of nebula, more ancient, more primordial, in order to envisage what the world was like in its initial state, and to finally respond to the impasse in which Descartes and Newton had left us. This primordial nebula is not made up of a cluster of stars, but of a cluster of particles that are at the origin of stellar and planetary systems. This is how the moment of cosmogenesis properly begins.

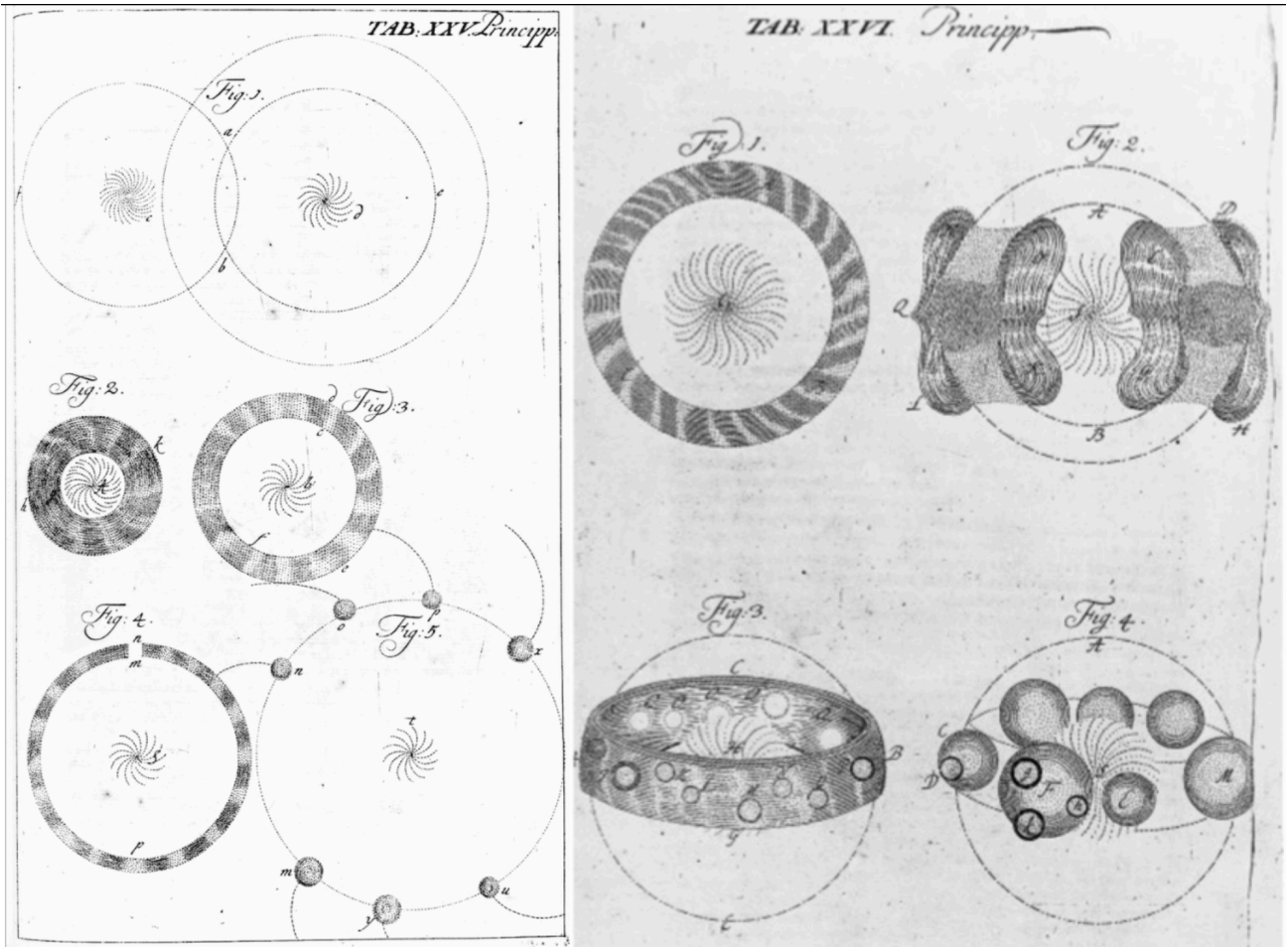

*Planetary evolution*. A series of diagrams taken from Swedenborg's *Principia* showing the general features of his theory.

The definition of these primitive particles that form a cloud of dust should lead us to ask : what theory of matter Kant is working from, in this text. In the preface, he had taken for himself a quotation that the Enlightenment had was in the habit of using to refer to the Cartesian cosmological gesture: « Give me matter and I will make a world of it ». However, the *Theory of the Heaven* cannot be said to pose the question of the essence of matter, as the *Physical monadology* published the following year will do. It's all in the formula: « « Give me matter and I will make a world of it » —> matter is given at the beginning. But even if the study of cosmology is not the study of particle physics, it cannot fail to produce differentiation in the matter with which it begins, in order to break up the uniformity. (So any cosmology will have even a minimal theory of the elements to introduce symmetry breaks and provoke motion. Indeed, if all the elements in the universe were perfectly equal, there would be nothing to suggest that there is change, and the cloud of particles would not be induced to transform itself into systems. And I think there is a hint here of the sense in which matter must be taken in order to make a cosmology.)

For Kant, if there are no qualitative differences between the elements as in ancient cosmologies between water, air, earth, fire... there are indeed different « kinds of elements » that are quantitatively different. Does this mean that some are bigger than others, as this was the case in Descarte's theory ? No, because Kant refused to reduce the essence of matter to some extension. On this point, he was closer to Leibniz (as can be seen from his early work on living forces). Above all, on this subject, Kant again took note of modern science, in particular of the advances in chemistry, which had applied the concept of density to the molecules of matter. So, in the *Theory of the Heavens,* the differences between particles will be differences in density.

At the beginning of the system, in a dust nebula, some molecules of matter are denser than others, and the laws are deeply rooted in it. Of course, Kant wanted these laws to be Newtonian, so the first major force running through the world were attraction. But he added a second force: repulsion, which must prevent all bodies from coming together in a single central point.

And from this conception of matter and these only two laws, Kant drew the model of the primordial nebula at the origin of all stellar systems: in such conditions, rest lasts only a moment, and the densest and lightest particles must have attracted the others to form a star. (You can see it on the screen in this image made by Swedenborg). Around it, the resulting and more massive matter must first have formed a disc of action, like in the rings of Saturn, before gathering into isolated masses that formed the planets, satellites and comets, which gravitate as a result of the cumulative forces of attraction and repulsion. And such a movement, born of the properties of matter in the solid, will continue its course in a space <u>now empty</u> because of inertia of the bodies.

This theory would have to be mathematically clarified, and in particular the role of repulsion, which is not consistent with Newtonian theory. This would be the role of Laplace. But the idea of a primitive nebula at the origin of stellar and planetary systems remains our current model.